\documentclass[%
reprint,
superscriptaddress,
showpacs,
amsmath,amssymb,
aps,
pra,
]{revtex4-1}

\usepackage{amsmath}
\usepackage{graphicx}
\usepackage{braket}
\usepackage{amssymb}
\usepackage{float}
\usepackage{color}
\usepackage[colorlinks=true,linkcolor=blue,anchorcolor=blue,citecolor=blue,urlcolor=blue]{hyperref}
\begin{document}
\preprint{APS/123-QED}

\title{Attosecond-Resolved Photoionization Dynamics and Interference-Enhanced Photoelectron Circular Dichroism in Chiral Molecules}

\author{Zheming Zhou}
\affiliation{School of Physics and Wuhan National  Laboratory for Optoelectronics, Huazhong University of Science and Technology, Wuhan 430074, China}
\author{Yang Li}
\email{liyang22@sjtu.edu.cn}
\affiliation{Key Laboratory for Laser Plasmas (Ministry of Education) and School of Physics and Astronomy, Collaborative Innovation Center for IFSA (CICIFSA), Shanghai Jiao Tong University, Shanghai 200240, China}
\author{Xu Zhang}
\affiliation{School of Physics and Wuhan National  Laboratory for Optoelectronics, Huazhong University of Science and Technology, Wuhan 430074, China}
\author{Yueming Zhou}
\email{zhouymhust@hust.edu.cn}
\affiliation{School of Physics and Wuhan National  Laboratory for Optoelectronics, Huazhong University of Science and Technology, Wuhan 430074, China}
\author{Peixiang Lu}
\email{lupeixiang@hust.edu.cn}
\affiliation{School of Physics and Wuhan National  Laboratory for Optoelectronics, Huazhong University of Science and Technology, Wuhan 430074, China}

\date{\today}

\begin{abstract}
Chiral molecules exhibit enantiosensitive light-matter interactions, with photoelectron circular dichroism (PECD) serving as a sensitive probe of molecular chirality through the asymmetry in the photoelectron wavepacket amplitude. Here, we demonstrate a photoelectron interferometric approach to access the phase of the photoelectron wavepacket and uncover attosecond dynamics in chiral molecule photoionization. Using circularly polarized attosecond XUV pulse trains synchronized with IR fields, we reveal distinct time delays between forward- and backward-ejected photoelectrons in a randomly oriented ensemble of chiral molecules. Moreover, we predict a pronounced enhancement of PECD due to the interference of the two photoionization pathways. The forward-backward time delay difference and the PECD are more prominent when the IR field counter-rotates with the XUV field. These results imply the counter-rotating IR field is more efficient in generating odd-parity photoelectron wavepackets in continuum-continuum transitions, highlighting the critical role of long-range chiral potential. Our work demonstrates a way of coherent control over the chiral photoelectron
wavepackets, providing a route to enhance chiral signals and manipulate ultrafast chiral dynamics on attosecond time scales.
\end{abstract}

\maketitle

Chirality, the geometric property of non-superimposable mirror-image structures, is a fundamental characteristic of a broad range of molecular systems with profound implications in chemistry, physics, and biology. In light-matter interactions, chiral molecules exhibit enantiosensitive responses, most notably photoelectron circular dichroism (PECD), which manifests as forward-backward asymmetries in photoelectron angular distributions upon ionization with circularly polarized light. While traditional probes of molecular handedness rely on weak magnetic-dipole effects, such as circular dichroism~\cite{CD}, PECD leverages electric-dipole-driven asymmetries in photoelectron angular distributions~\cite{PECD1}, yielding chiral signals that can reach several percent. Since its first experimental realization~\cite{PECD2}, PECD has been observed across various ionization regimes, from one-photon~\cite{one1,one2,one3,one4,one5,one6} and multiphoton~\cite{multi1,multi2,multi3,multi4,multi5,multi6,multi7} to strong-field~\cite{strong1,strong2,strong3,strong4,strong5,strong6} ionization processes. 

Despite revolutionizing gas-phase chiral recognition, PECD provides only a partial view of photoionization dynamics. It primarily captures amplitude asymmetries of photoelectron wave packets (EWPs) while leaving their phase, a critical determinant of electron dynamics, largely unresolved. The retrieval of the EWP phase is essential for a comprehensive understanding of chiral photoionization. Electron interferometry offers a powerful means to extract phase information. Recent advances in attosecond photoelectron interferometry \cite{interferometry1,interferometry2} have facilitated phase-sensitive measurements of EWPs in randomly oriented chiral molecular ensembles. However, in the strong-field regime, the complexity arising from multiphoton ionization pathways and the dependence of phase on the driving field intensity pose significant challenges in interpreting these measurements. These difficulties hinder the characterization of chiral photoionization and, consequently, the extraction of fundamental chiral properties of molecular systems. The reconstruction of attosecond beating by interference of two-photon transitions (RABBITT)~\cite{RABBITT} has emerged as a cornerstone technique for measuring attosecond-scale photoelectron time delays in atoms~\cite{atomicRABBITT1,atomicRABBITT2,atomicRABBITT3,atomicRABBITT4,atomicRABBITT5,atomicRABBITT6} and molecules~\cite{molecularRABBITT1,molecularRABBITT2,molecularRABBITT3,molecularRABBITT4,molecularRABBITT5,molecularRABBITT6}. By exploiting weak laser fields, RABBITT delivers intensity-independent delays, providing direct access to inherent molecular properties. Recent advances in generating circularly polarized extreme ultraviolet (XUV) attosecond pulses~\cite{CPAPT1,CPAPT2,CPAPT3,CPAPT4,CPAPT5} now enable the application of RABBITT to chiral systems, opening a route to probe enantiosensitive phase dynamics in photoionization.

In this Letter, we demonstrate a RABBITT scheme combining circularly polarized XUV attosecond pulse trains and IR fields to resolve attosecond electron dynamics in two-photon ionization of randomly oriented chiral molecules. From the angle-resolved photoelectron spectra, we analyze both the amplitude and phase of the ionized EWP, i.e., PECD and photoionization time delays. Our results reveal significantly enhanced PECD magnitudes compared to those in the previous single-photon ionization case, due to the interference of the two-photon-transition pathways in the RABBITT scheme. Moreover, we unveil a distinct difference in photoionization time delays between forward- and backward-emitted photoelectrons. Notably, the differential time delay signal is amplified in counter-rotating XUV-IR configurations, highlighting the critical role of continuum-continuum transitions in enantiosensitivity. 

\begin{figure}[t]
	\includegraphics[width=8.7cm]{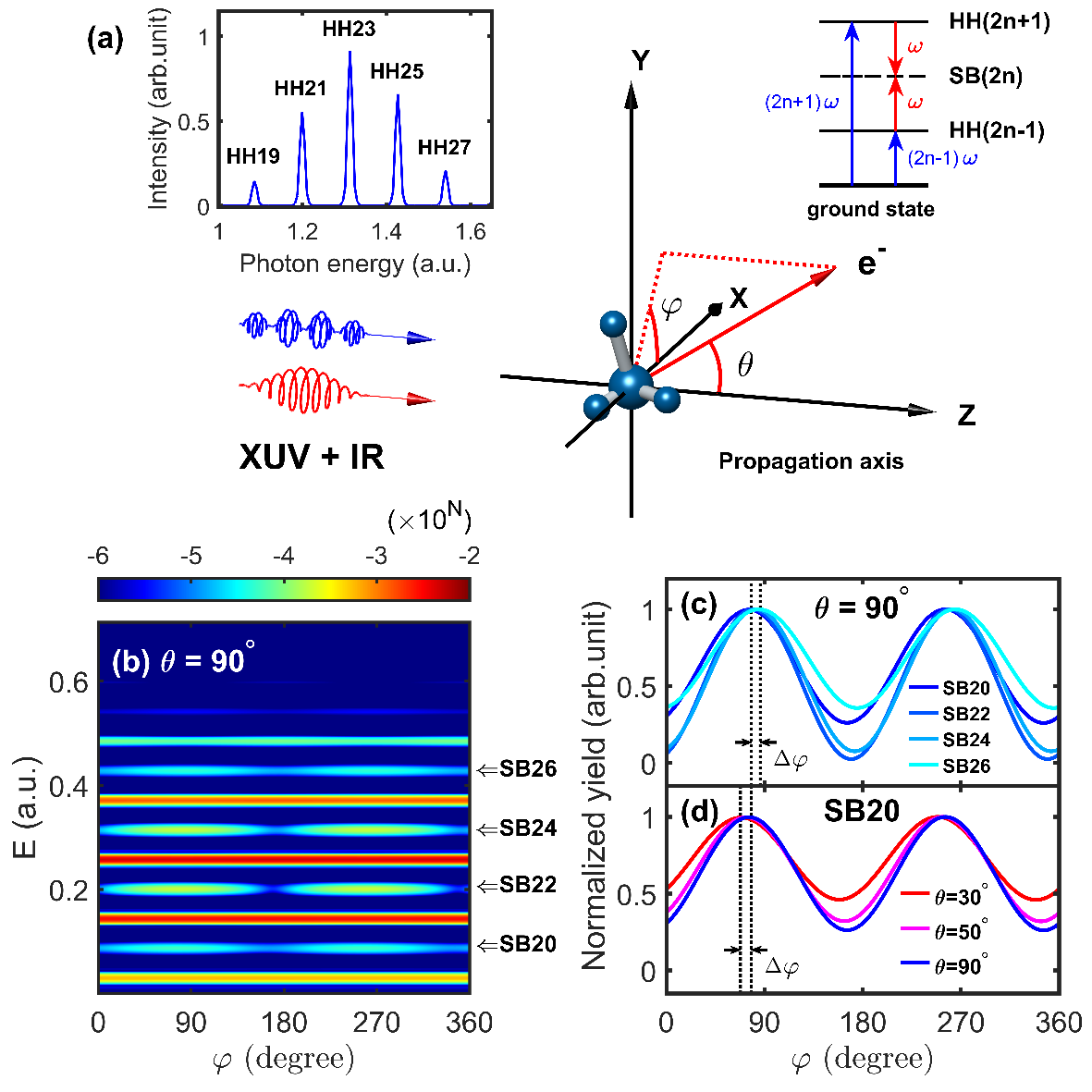}
	\caption{\label{F1}
        (a) Scheme for detecting chiral molecules by circularly polarized RABBITT. The laser pulses propagate along the z-axis. The ground state electrons first transition to the intermediate continuum states (main peaks) by absorbing an XUV photon with different energy, then transition to the sidebands by absorbing or emitting an IR photon to form the interference fringes. (b) Photoelectron angular distribution in the laser polarization plane ($\theta=90^{\circ}$). The peak intensities of XUV and IR fields are $1\times10^{12}\,\mathrm{W/cm^{2}}$ and $1\times10^{11}\,\mathrm{W/cm^{2}}$, respectively, and the wavelength of IR field is 800 nm (c) $\varphi$-dependent photoelectron yields of different SBs in the polarization plane. (d) $\varphi$-dependent photoelectron yields of SB20 at different emission angles.
	}
\end{figure}

To this end, we perform quantum-mechanical calculations by solving time-dependent Sch\"{o}rdinger equation (TDSE). Employing the velocity gauge, we calculate the photoelectron momentum distribution of a model chiral molecule~\cite{multi2} in dipole approximation. The effective chiral potential, $V({\textbf{r}})=\sum_{i=1}^{4}-Z_{i}/|\textbf{r}-\textbf{R}_{i}|$, consists of four nuclei positioned at $\textbf{R}_{1}=\textbf{0}$, $\textbf{R}_{2}=\hat{\textbf{x}}$, $\textbf{R}_{3}=2\hat{\textbf{y}}$ and $\textbf{R}_{4}=3\hat{\textbf{z}}$, respectively, with charges  $Z_{1}=-5.0$ a.u. and $Z_{2-4}=2.0$ a.u. Utilizing the single-center method~\cite{SC1,SC2,SC3}, we expand the electron wavefunction and chiral potential in terms of spherical harmonics $Y_{lm}(\theta,\varphi)$.
The radial coordinate is discretized using the finite-element discrete variable representation~\cite{FEDVR1,FEDVR2,FEDVR3}. The wavefunction is propagated using the split-Lanczos method~\cite{Lanczos}. The ionized electron wavefunction is captured through an absorption edge, and the photoelectron momentum distribution is obtained by projecting the wavefunction onto continuum states. For randomly oriented chiral molecules, we calculate the orientation-averaged momentum distribution~\cite{Djiokap}
\begin{equation}
    I(\textbf{k})=\int d\hat{\textbf{R}}\,|\psi(\hat{\textbf{R}};\textbf{k})|^{2},
    \label{E6}
\end{equation}
where $\psi(\hat{\textbf{R}};\textbf{k})$ is the ionization amplitude for a fixed molecular orientation $\hat{\textbf{R}}$, and $\textbf{k}$ denotes the momentum in the laboratory frame. In our calculations, the integral is evaluated by numerical quadrature over discretized molecular orientations with Euler angular spacing $\Delta\alpha=\Delta\beta=\Delta\gamma=\pi/6$. The convergence has been verified by decreasing angular spacings.

\begin{figure*}[t]
    \includegraphics[width=17cm]{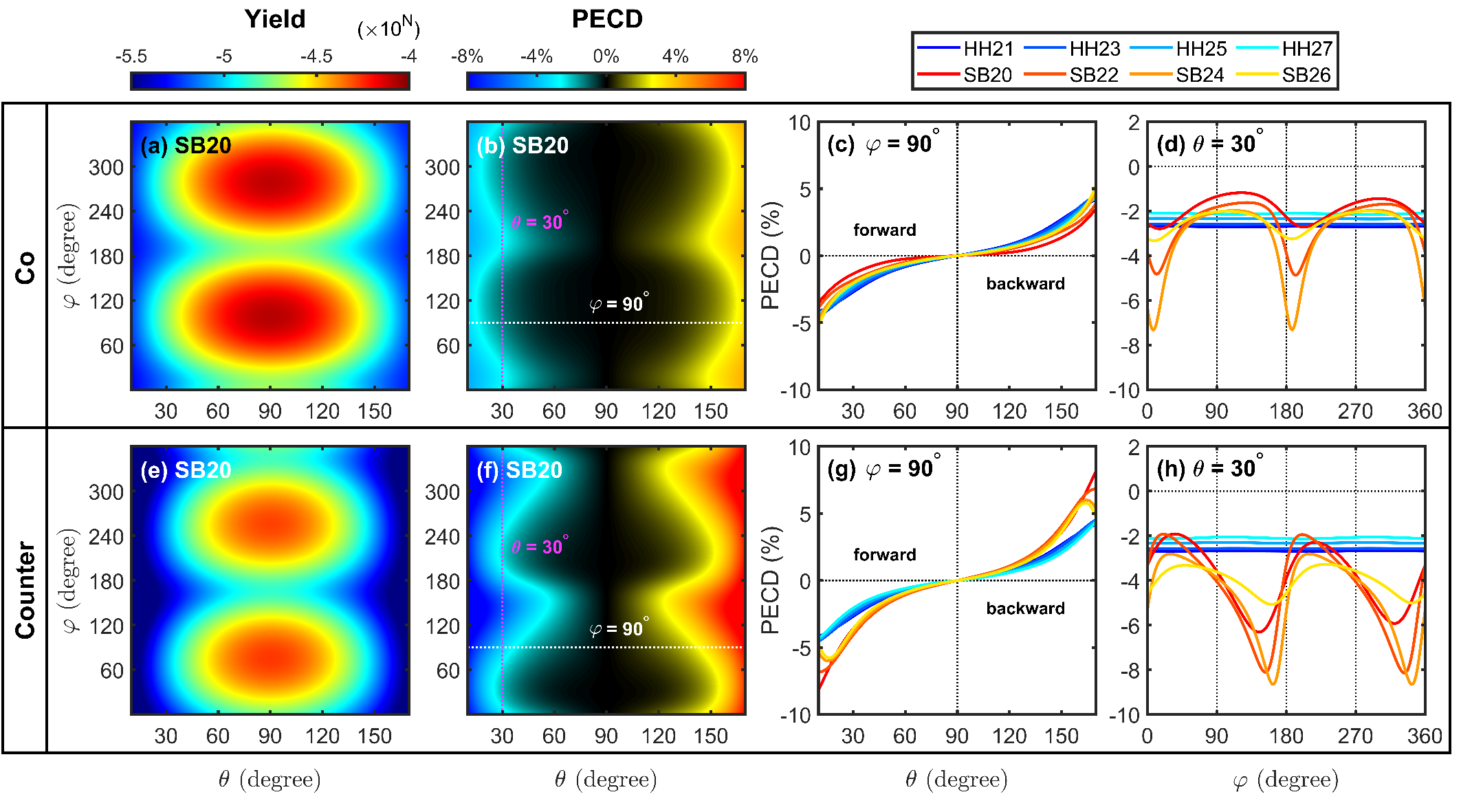}
    \caption{\label{F2}
        (a) Angle-resolved photoelectron distribution for SB20. (b) Angle-resolved PECD for SB20. (c) The PECD as a function of $\theta$ at $\varphi=90^{\circ}$ for different main peaks and SBs. (d) The PECD as a function of $\varphi$ when $\theta=30^{\circ}$ for different main peaks and SBs. The XUV and IR fields are co-rotating circularly polarized. (e)-(h) The same as (a)-(d) but for the counter-rotating fields.
        }
\end{figure*}

Figure~\ref{F1}(a) illustrates the laser configuration and the coordinate system adopted in this work. The circularly polarized attosecond XUV pulse train comprises odd-order harmonics from HH19 ($29.42\,\mathrm{eV}$) to HH27 ($41.80\,\mathrm{eV}$). The synchronized 800-nm IR field is tuned to be circularly polarized with the same or opposite helicities with respect to the XUV pulse. The two fields are co-polarized in the $x$-$y$ plane and propagate along the $z$ axis. In the standard RABBITT scheme with linearly polarized light, sideband (SB) oscillations as functions of the XUV-IR time delay are recorded, and the corresponding photoionization time delay is obtained by analyzing these RABBITT traces. In circular RABBITT for the unoriented molecule ensemble, the photoionization time delay can be determined via $\varphi$-resolved photoelectron angular distribution without time-delay scanning~\cite{SM}. This property greatly reduces the computational cost, which is important for the theoretical study of chiral molecules.

Figure~\ref{F1}(b) shows the calculated $\varphi$-resolved photoelectron energy spectrum in the polarization plane for counter-rotating XUV and IR fields. The five main peaks correspond to photoionization by HH19 to HH27, and the four SBs record the interference of absorption and emission pathways where one IR photon is absorbed (emitted) from the lower(higher)-energy main peak. The distributions of the main peaks are $\varphi$-independent due to the perturbative nature of the IR field. The SB signals exhibit $2\varphi$-oscillations. Figures~\ref{F1}(c) and~\ref{F1}(d) show photoelectron yields for different SBs in the polarization plane ($\theta=90^\circ$), and for SB20 at varying polar angles $\theta$, respectively. In both cases, clear phase shifts of the RABBITT traces as functions of azimuthal angle $\varphi$ exist, which encode the phase information of the EWPs and provide rich information of chiral molecule photoionization, as demonstrated below.

Before inspecting the phase information, we first analyze the amplitudes of the EWPs. Figures~\ref{F2}(a) and~\ref{F2}(e) show the photoelectron angular distributions for SB20 by co- and counter-rotating laser fields, respectively. The photoelectron yields of the SBs in the co-rotating fields are several times higher than those of the counter-rotating fields. The corresponding distributions of PECD, defined as $\text{PECD}(\theta,\varphi)=[I(\theta,\varphi)-I(\pi-\theta,\varphi)]/[I(\theta,\varphi)+I(\pi-\theta,\varphi)]$, where $I(\theta,\varphi)$ denotes the photoelectron angular distributions, are shown in Figs.~\ref{F2}(b) and~\ref{F2}(f). The PECD depends on the azimuthal angle $\varphi$, and reaches values up to $3\%$ and $8\%$ for the co-rotating and counter-rotating fields, respectively. Slices of PECD for different SBs at $\varphi=90^\circ$ [Figs.~\ref{F2}(c) and~\ref{F2}(g)] and $\theta=30^{\circ}$ [Figs.~\ref{F2}(d) and~\ref{F2}(h)] highlight its angular dependence. The PECD signals of main peaks are also presented for comparison. For both the main peaks and SBs, the PECD signals increase with the polar angle $\theta$. At $\theta=30^{\circ}$, SB PECD signals show irregular oscillation with $\varphi$, while the main peak PECDs remain isotropic. The maximum value of PECD signals in the SBs exceeds 8\%, significantly surpassing those of the main peaks. Moreover, the PECD signals of sidebands in the counter-rotating fields are generally larger than the co-rotating fields. 

To understand the enhanced PECD signal and its irregular oscillation with $\varphi$, we generalize the theory of single- and two-photon ionization of chiral molecules~\cite{theory1,theory2,theory3,theory4} to circular RABBITT, incorporating interference of the absorption and emission two-photon pathways~\cite{SM}. The photoelectron angular distribution of the randomly oriented chiral molecular ensemble at SBs can be expanded in terms of spherical harmonics~\cite{SM},
\begin{equation}
    \begin{aligned}
        I(\theta,\varphi)=\sum_{m=0,\pm2}\sum_{l=|m|}^{4}\beta_{lm}Y_{lm}(\theta,\varphi),
    \end{aligned}
    \label{E1}
\end{equation}
where the anisotropy parameters $\beta_{lm}$ are generally complex numbers $\beta_{lm}=|\beta_{lm}|e^{i\delta_{lm}}$ and $\beta_{l,-m}=\beta_{lm}^{*}$. $\beta_{l0}$ corresponds to the individual contribution of the two-photon absorption and emission pathways, and forms the background signal in the PECD of the SBs. The contributions of $\beta_{l,\pm2}$ result from the interference of the absorption and emission pathways~\cite{SM}, 
and they are responsible for the $2\varphi$-modulation of SB signals in Figs.~\ref{F1}(c) and ~\ref{F1}(d). The resulting SB-PECD can be expressed as~\cite{SM}
\begin{widetext}
\begin{equation}
        \text{PECD}(\theta,\varphi)=\frac{\sum\limits_{l=1,3}\beta_{l0}\text{A}_{l0}\text{P}_{l}(\cos\theta)
        +2\,|\beta_{32}|\text{A}_{32}\text{P}_{3}^{2}(\cos\theta)\cos(2\varphi+\delta_{32})}
        {\sum\limits_{l=0,2,4}\beta_{l0}\text{A}_{l0}\text{P}_{l}(\cos\theta)
        +2\sum\limits_{l=2,4}|\beta_{l2}|\text{A}_{l2}\text{P}_{l}^{2}(\cos\theta)\cos(2\varphi+\delta_{l2})},
    \label{PECD}
\end{equation}
\end{widetext}
where $\text{A}_{lm}$ accounts for the normalized coefficient of the associated Legendre polynomials. As shown in Eq.~\eqref{PECD}, anisotropy parameters with odd $l$, i.e., $\beta_{10}$, $\beta_{30}$, $\beta_{3,\pm2}$, are responsible for the PECD signal. The first two terms ($m=0$) in the numerator correspond to the PECD from the incoherent sum of the two pathways, which are independent of $\varphi$. The third term ($m=2$) is due to the interference of the two pathways, resulting in the $\varphi$ modulation of the SB PECDs in Fig.~\ref{F2}. Interference terms also appear in the denominator, and thus the PECD shows irregular oscillation with $\varphi$ in Figs.~\ref{F2}(d) and~\ref{F2}(h). These results indicate that PECD in chiral molecules can be greatly enhanced through interference.

\begin{figure}[t]
    \includegraphics[width=8.5cm]{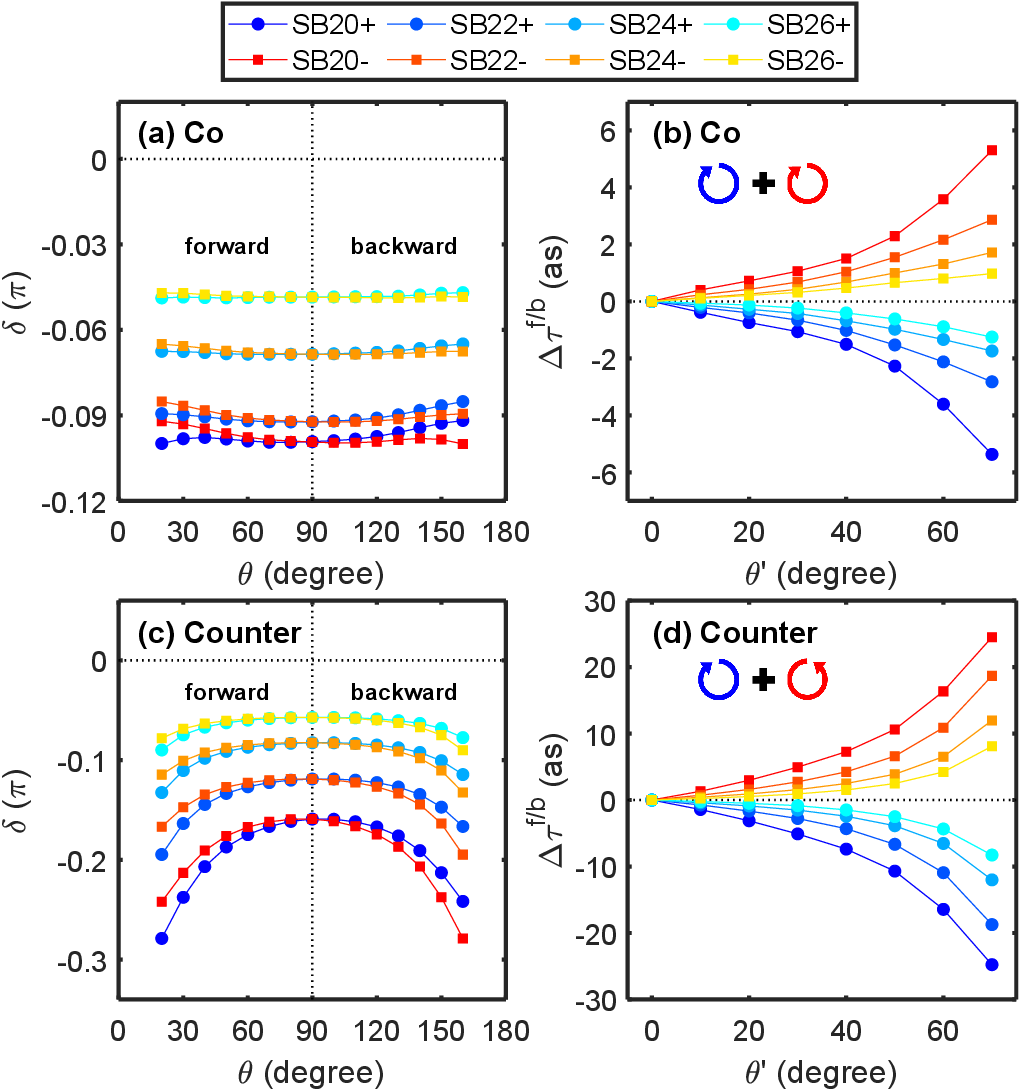}
    \caption{\label{F3}
        (a) RABBITT phase and (b) differential delay for co-rotating circularly polarized fields. 
        The results of two enantiomers (denoted as + and -, respectively) are represented by cold and warm color lines for comparison. (c) (d) The same as (a) and (b) but for the counter-rotating circularly polarized fields. Here $\theta'=90^{\circ}-\theta$ is the emission angle with respect to the polarization plane.
        }
\end{figure}

Now, we turn to the photoionization time delay, which directly reflects the phase of the photoelectrons. We obtain this information by analyzing the 2$\varphi$-oscillation of the photoelectron yield in SBs [Figs.~\ref{F1}(c) and~\ref{F1}(d) ]. For this purpose, we rewrite the photoelectron angular distribution of SBs [Eq.~\eqref{E1}] to show the oscillation explicitly,

\begin{equation}
    \begin{aligned}
        I(\theta,\varphi)=A(\theta)-B(\theta)\cos\left[2\varphi-\frac{\delta(\theta)}{\mathcal{M}}\right].
    \end{aligned}
    \label{E2}
\end{equation}
Here, $\delta(\theta)$ represents the RABBITT phase and the photoionization time delay is $\tau(\theta)=\delta(\theta)/2\omega$, where $\omega$ is the angular frequency of the IR field. The parameter $\mathcal{M}=\pm1$ indicates the opposite (counter-rotating) or same (co-rotating) helicity of the IR field with respect to the XUV field. We define the forward-backward differential time delay $\Delta\tau^{f/b}=\tau^{f}-\tau^{b}$~\cite{interferometry1}, as a measure of the phase-based chiral response. Figures~\ref{F3}(a) and~\ref{F3}(c)  show the extracted RABBITT phases for two enantiomers in co- and counter-rotating circularly polarized fields, respectively. 
Different from the atomic system~\cite{atomicRABBITT4,atomicRABBITT5,atomicsystem1,atomicsystem2}, an obvious forward-backward asymmetry in the phase is observed. This asymmetry is reversed for the two enantiomers. Figs.~\ref{F3}(b) and~\ref{F3}(d) display the differential time delay, and the nonzero value directly reflects the chiral character of the molecular potential on the leaving photoelectron. The absolute value increases with the emission angle $\theta'$ ($\theta'=\frac{\pi}{2}-\theta$). For the lower-order SBs, the differential time delay is larger because lower-energy photoelectrons experience a stronger influence from the chiral potential. 

\begin{figure}[t]
    \includegraphics[width=9cm]{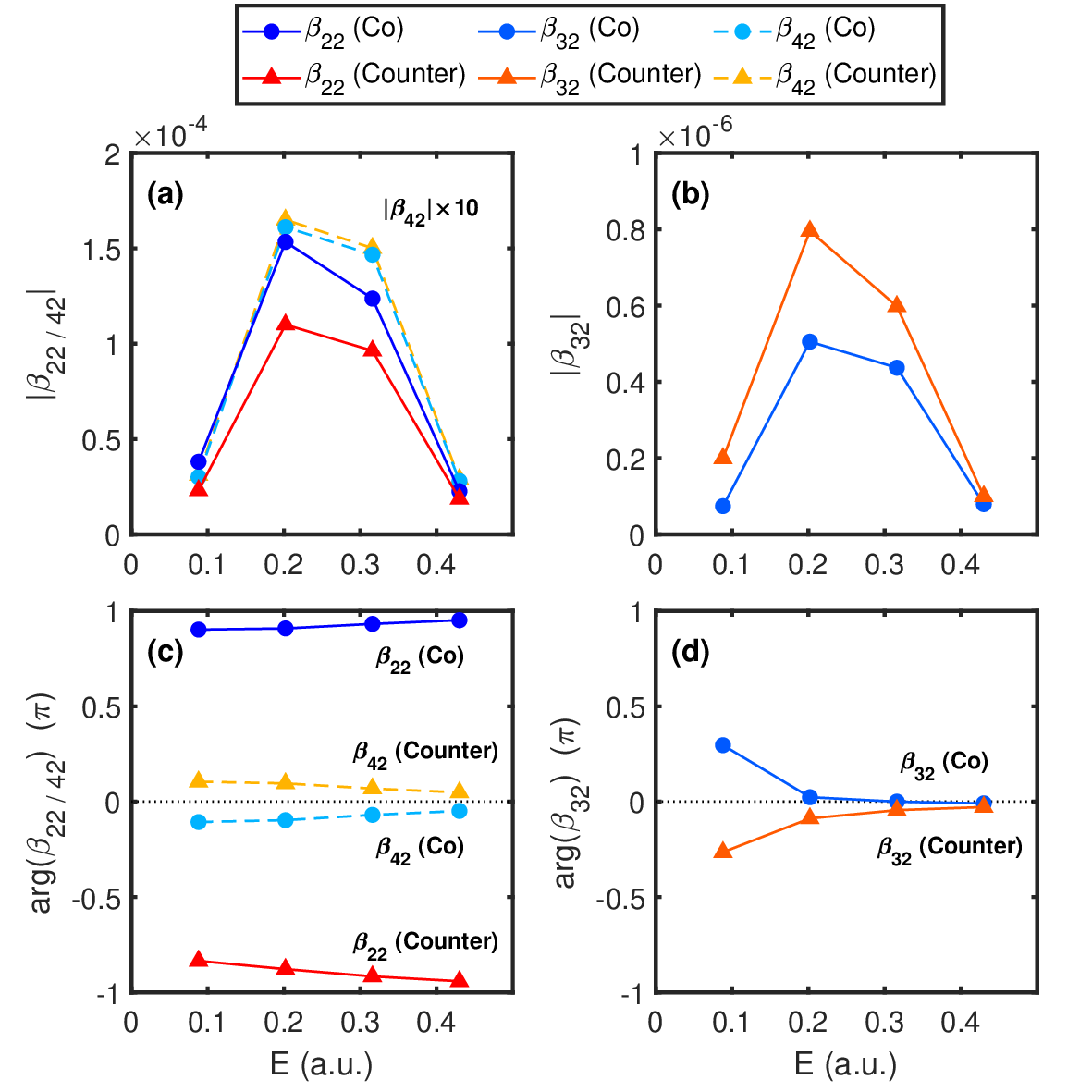}
    \caption{\label{F4}
        Anisotropy parameters for enantiomer(+) in co- and counter-rotating fields. (a) Amplitudes of $\beta_{22}$ and $\beta_{42}$. Here, $\beta_{42}$ has been magnified by a factor of 10. (b) Amplitude of $\beta_{32}$. (c) Phases of $\beta_{22}$ and $\beta_{42}$. (d) Phase of $\beta_{32}$. The amplitudes and phases of $\beta_{22}$, $\beta_{32}$, $\beta_{42}$ are denoted by circles, triangles, and squares, respectively. The cold and warm color lines respectively represent the cases of co-rotating and counter-rotating fields.
        }
\end{figure}

Interestingly, the differential time delays observed in the counter-rotating fields are much larger than those in the co-rotating fields. For the present model molecule, it reaches up to 25 attoseconds in the counter-rotating fields. This behavior highlights the sensitivity of the time delay to the chirality-induced continuum-continuum (CC) transitions. It is in contrast to the previous study in strong-field multiphoton ionization of chiral molecules, where the differential delay induced by CC transition is found to be negligible ~\cite{interferometry1}, probably due to the CC transitions being insensitive to the chiral character of the long-range molecular potential in the strong-field regime. For our scheme in the weak-field region, however, the differential time delay directly reflects the intrinsic properties of the chiral system, establishing CC transition as a highly sensitive tool to probe the long-range chiral interactions. 

Deeper insights about the difference in the differential delay between the co- and counter-rotating fields can be obtained by decomposing the RABBITT phase into a series of associated Legendre polynomials~\cite{SM}
\begin{equation}
    \begin{aligned}
       \delta(\theta)=-\arg\left[-\sum_{l=2}^{4}\beta_{l2}\rm{A}_{\mathit{l}2}\rm{P}_{\mathit{l}}^{2}(\cos\theta)\right].
    \end{aligned}
   \label{E5}
\end{equation}
The amplitudes and phases of these anisotropy parameters $\beta_{l2}$ are obtained by projecting the photoelectron angular distribution of the SBs on the corresponding spherical harmonic functions, as shown in Fig.~\ref{F4}. The differential time delay can be approximated as~\cite{SM}

\begin{equation}
    \begin{aligned}
        \Delta\tau^{f/b}
        \approx\frac{\mathcal{M}\sin(\delta_{22}-\delta_{32})\,|\beta_{32}\rm{A}_{32}\rm{P}_{3}^{2}(\sin\theta')|}
        {\omega\,|\beta_{22}\rm{A}_{22}\rm{P}_{2}^{2}(\sin\theta')+\beta_{42}\rm{A}_{42}\rm{P}_{4}^{2}(\sin\theta')|}.
    \end{aligned}
    \label{E9}
\end{equation}
The anisotropy parameter $\beta_{32}$ governs the forward-backward asymmetry of the time delay. For an achiral molecule ensemble, $\beta_{32}$ is strictly zero after the molecular orientation average~\cite{SM}, resulting in a vanishing forward-backward differential time delay. The remaining two anisotropy parameters $\beta_{22}$ and $\beta_{42}$ in the denominator influence the overall magnitude of the differential time delay. As presented in Fig.~\ref{F4}, $\beta_{42}$ is much smaller than $\beta_{22}$ and is insensitive to the helicity of the IR field. In contrast, both $\beta_{22}$ and $\beta_{32}$ exhibit a clear dependence on the helicity of the IR field. In counter-rotating fields, the amplitude of $\beta_{32}$ is larger while $\beta_{22}$ is smaller, indicating the counter-rotating IR field is more efficient in generating odd-parity EWPs. Moreover, the phase difference between $\beta_{22}$ and $\beta_{32}$, i.e., $\delta_{22}-\delta_{32}$, is almost independent of the helicity of the IR field. As a consequence, $|\Delta\tau^{f/b}_{\rm{Co}}|<|\Delta\tau^{f/b}_{\rm{Counter}}|$, as shown in Fig.~\ref{F3}.

In conclusion, we have demonstrated an interferometric approach based on two-photon ionization to study the attosecond chiral dynamics. Our results revealed a significant enhancement in PECD, arising from the interference between two ionization pathways. Moreover, we unveiled a remarkable asymmetry in the photoionization time delay between forward and backward-emitted photoelectrons. Both the PECD enhancement and the time delay asymmetry are much more prominent in the counter-rotating fields. These features highlight the critical role of molecular long-range potentials in chiral response, which can be probed with CC transitions. With the advanced attosecond light sources~\cite{CPAPT4,CPAPT5,atomicsystem1,atomicsystem2}, all these findings are experimentally feasible. The emerging techniques such as three-sideband RABBITT~\cite{CC1, CC2} could be applied to resolve the phase dynamics of CC transitions in chiral systems. 
Notably, we demonstrate coherent control over the parity of photoelectron wavepackets via the helicity of the infrared field, providing a route to enhance chiral signals and manipulate ultrafast chiral dynamics on attosecond time scales.

\textit{Acknowledgments}\textemdash This work was supported by National Key Research and Development Program of China (Grant No. 2023YFA1406800), National Natural Science Foundation of China (Grants No. 12374264, 12274294, 12021004, 12204545, 12434010), Basic Research Support Program of Huazhong University of Science and Technology (2024BRA002). The computing work in this paper is supported by the Public Service Platform of High Performance Computing provided by Network and Computing Center of HUST.

Z. Zhou and Y. Li contributed equally to this work.

\end{document}